\documentclass[aip,jcp,reprint]{revtex4-2}

\usepackage[T1]{fontenc}
\usepackage{graphicx,amsmath,amssymb,bm}

\newcommand{\mx}{_\mathrm{x}}
\newcommand{\mc}{_\mathrm{c}}
\newcommand{\mxc}{_\mathrm{xc}}
\newcommand{\mks}{_\mathrm{KS}}
\newcommand{\rs}{r_\mathrm{s}}
\newcommand{\kf}{k_\mathrm{F}}
\newcommand{\ks}{k_\mathrm{s}}
\newcommand{\br}{\bm{r}}
\newcommand{\xks}{\chi\mks}
\newcommand{\vc}{v_\mathrm{c}}
\newcommand{\et}{\widetilde{\epsilon}}
\newcommand{\ext}{_\mathrm{ext}}

\begin{document}

\title{Revealing quasi-excitations in the low-density homogeneous electron gas with model exchange-correlation kernels}

\author{Aaron D. Kaplan}
\email{adkaplan@lbl.gov}
\affiliation{Materials Project, Lawrence Berkeley National Laboratory, Berkeley, CA 94720}

\author{Adrienn Ruzsinszky}
\affiliation{Department of Physics and Engineering Physics, Tulane University, New Orleans, LA 70118}

\date{\today}

\begin{abstract}
    Time-dependent density functional theory (TDDFT) within the linear response regime provides a solid mathematical framework to capture excitations. 
    The accuracy of the theory, however, largely depends on the approximations for the exchange-correlation (xc) kernels. 
    Away from the long-wavelength (or $q=0$ short wave-vector) and zero-frequency ($\omega=0$) limit, the correlation contribution to the kernel becomes more relevant and dominant over exchange. 
    The dielectric function in principle can encompass xc effects relevant to describe low-density physics. 
    Furthermore, besides collective plasmon excitations, the dielectric function can reveal collective electron-hole excitations, often dubbed ``ghost excitons.''
    Beside collective excitons, the physics in the low-density regime is rich, as exemplified by a static charge-density wave that was recently found for $\rs > 69$, and was shown to be associated with softening of the plasmon mode.
    These excitations are seen to be present in much higher density 2D HEGs, of $\rs \gtrsim 4$.
    In this work we perform a thorough analysis with xc model kernels for excitations of various nature. 
    The uniform electron gas, as a useful model of real metallic systems, is used as a platform for our analysis. 
    We highlight the relevance of exact constraints as we display and explain screening and excitations in the low-density region.
\end{abstract}

\maketitle

\section{Introduction}

The homogeneous electron gas (HEG) \cite{giuliani2005} underlies practical approximations in density functional theory (DFT) \cite{hohenberg1964,kohn1965} and is a useful model of metallic systems.
Density functional approximations built upon the constraint of the HEG limit have predictive power to capture a wide range of electronic phenomena.
The same predictive power transfers to excitations in the HEG from time-dependent DFT (TDDFT) \cite{runge1984,gross1985, fundamentalsTDDFT}.
The interacting HEG, sometimes called jellium, is an important model because it has a Hamiltonian with Coulomb repulsion between electrons, but with an external potential that arises from a uniform positive-charge background.
The HEG total energy minimizes at Wigner-Seitz radius $\rs \approx 4$ bohr, mimicking the valence electron density of metallic sodium.   

In linear response TDDFT \cite{petersilka1996, bottiTDDFT, maitraTDDFT}, the exact linear density-density response function $\chi(\br,\br',\omega)$ of an electronic system in its ground state to a weak, frequency or $\omega$-dependent external scalar potential $\delta v(\br',\omega)$ delivers the exact excited-state energies $\hbar \omega_0$ of the system when $\chi^{-1}(\br,\br',\omega_0)=0$.
The interacting density-density response function $\chi$ contains the poles of the real system, while the non-interacting or Kohn-Sham response function $\xks$ contains the poles of the fictitious Kohn-Sham system.
For the HEG, $\xks$ is the Lindhard function \cite{lindhard1954}.

To boost the accuracy of $\xks$, xc effects can be added to the bare Coulomb interaction $\vc(q) = 4\pi/q^2$ in the form of xc kernels $f\mxc(q,\omega)$.
Kernels in principle are rigorous derivatives of the xc potential, and are often modeled by satisfying exact constraints \cite{lein2000, maitraTDDFT}.
xc effects then screen $\xks$
\begin{equation}
    \chi(q,\omega) = \xks(q,\omega) \et^{-1}(q,\omega),
    \label{eq:chi_inter}
\end{equation}
through the wave-vector $q$ and frequency $\omega$-dependent dielectric function
\begin{equation}
    \et(q,\omega) = 1 - \left[\vc(q) + f\mxc(q,\omega) \right] \xks(q,\omega).
    \label{eq:DF_chi}
\end{equation}

Plasmons or collective electronic excitations in the interacting HEG \cite{pines1952,pines1963} are defined by those complex frequencies where $|\et| = 0$.
However, since long-range screening is assumed to be perfect in metals, excitons are not usually expected in metals \cite{giuliani2005,fetter1971}, including the HEG.

More accuracy is guaranteed when xc kernels are modeled by satisfying mathematical constraints in the same spirit as some xc density functional approximations are designed for the xc energy $E\mxc[n]$.
Unlike ultra-nonlocal kernels, which diverge as $\lim_{q\to 0} f\mxc \sim 1/q^2$, \cite{ghosez1997,bottiLR,bootstrap2011,ullrich2016}, short-range kernels of the HEG \cite{moroni1995,corradini1998,constantin2007,ruzsinszky2020,kaplan2022,kaplan2023}, where $\lim_{q \to 0} f\mxc \to \mathrm{const}$, do not yield bound excitons \cite{albrecht1998}.

The simplest xc kernel is the adiabatic local density approximation (ALDA)\cite{hohenberg1964,kohn1965} and is built upon the HEG paradigm.
It is spatially local, but at $q \to 0$, it satisfies the compressibility sum rule,
\begin{equation}
    \lim_{q \to 0} \left[\lim_{\omega \to 0} f\mxc(q,\omega) \right] = \frac{d^2}{dn^2} (n \varepsilon\mxc^\mathrm{HEG})\bigg|_{n = n(\br)},
\end{equation}
where $\varepsilon\mxc^\mathrm{HEG}$ is the exchange-correlation energy per electron in a spin-unpolarized HEG. 
ALDA is not expected to accurately reproduce excitations in the HEG, but it forms a basis for more sophisticated approximations to $f\mxc$.
That approach was taken in the work by Corradini \textit{et al.} \cite{corradini1998}.

\section{Static Screening and the Low-Density Regime of the HEG}

Screening is fundamental to understanding excitations in the HEG and real materials \cite{giuliani2005,fetter1971,knox1963,bassani1975,bechstedt2014,mahan1981}.
While screening is positive at high densities, the low-density regime of the HEG can feature negative screening which alters the nature of its collective excitations \cite{takada2005,takada2016,koskelo2023}.
Mathematically, negative screening reflects poles of the dielectric function $\epsilon$ on the imaginary frequency axis \cite{takayanagi1997}.

Within this work, we discuss (i) the improved description of screening, and (ii) predicted excitations in the low-density regime of the HEG found from recently-developed xc kernels within TDDFT \cite{ruzsinszky2020,kaplan2022,kaplan2023}.
These model kernels help us highlight the role of exact mathematical constraints for the low density HEG, or in general for metals.
Negative screening due to imaginary-frequency poles of the dielectric function are associated with collective ``ghost excitonic'' modes \cite{takayanagi1997}.

The zeroth-order approach in TDDFT is the random phase approximation (RPA) \cite{nozieres1958,langreth1975}, which sets $f\mxc^\mathrm{RPA} = 0$.
Since the region of collective excitations below a critical wave-vector is dominated by the Coulomb interaction, the RPA yields reasonable plasmon dispersion \cite{nepal2020}.
RPA is known for \emph{not} yielding bound excitons in semiconductors and insulators \cite{albrecht1998}.
Koskelo, Reining, and Gatti \cite{koskelo2023} recently demonstrated that excitons can exist even in metals.
This fact may not be surprising given imperfect screening in metals at short range.
As shown in real space, RPA can potentially capture excitons with a short electron-hole distance. 
With the RPA input dielectric function, the many-body GW-BSE (Bethe-Salpeter equation) should in principle exhibit excitons in metals \cite{martin2016,spataru2004}.
However, this is not the case for plasmon excitations because BSE inconsistently treats the input and output dielectric functions \cite{koskelo2023}.
Although many-body approximations are often not matched in accuracy by TDDFT, the inconsistency between the electron-hole interaction and self-energy in the former theory is often a bottleneck toward further improvement.

In this work, we motivate the use of approximate, physically-constrained kernels in TDDFT for exploring excitonic behavior in real materials.
Such an approach is computationally advantageous over GW-BSE.
To capture more than just qualitative physics at low density, one needs to go beyond ALDA.
The static HEG kernel $f\mxc(q,\omega=0)$, calculated via quantum Monte Carlo (QMC) \cite{moroni1995,kukkonen2021} and parameterized in Refs. \citenum{corradini1998} and \citenum{kaplan2023}, already improves upon the ALDA.

The authors have recently co-developed various increasingly-more accurate versions of a family of spatially- and temporally-non-local kernels called MCP07.
The MCP07 kernel \cite{ruzsinszky2020} is a constraint-based model of $f\mxc(q,\omega)$ that interpolates between the static limit of a modified Constantin-Pitarke 2007 (CP07) \cite{constantin2007} $f\mxc(q,\omega=0)$, and the Gross-Kohn-Iwamoto \cite{gross1985,iwamoto1987} dynamic LDA $f\mxc(q=0,\omega)$.
The changes in the static MCP07 kernel relative to CP07 affect both its $q \to 0$ and $q\to\infty$ limits.
Frequency-dependence is introduced to the static MCP07 kernel on a length scale that also controls the approach to the $f\mxc(q\to\infty,\omega=0)$ limit.
MCP07 has a broad range of successes from accurate correlation energies per electron for the HEG \cite{ruzsinszky2020,kaplan2022}, finite lifetimes for plasmons\cite{ruzsinszky2020} of small non-zero wavevector, to the right low-density behavior in which a static charge-density wave \cite{ruzsinszky2020,perdew1980,perdew2021} arises from a softening of the plasmon mode.

The static properties of MCP07 are superseded by the recently-developed Kaplan-Kukkonen kernel (AKCK) \cite{kaplan2023}.
AKCK analytically models both the static density $G_+(\rs,q,\omega=0)$ and spin $G_-(\rs,q,\omega=0)$ local field factors for the HEG, with free parameters determined by a fit to small-$q$ QMC data \cite{kukkonen2021}.
The density local-field factor is related to the xc kernel as $G_+ = -q^2 f\mxc/(4\pi)$.

The revised MPCP07 (rMCP07) kernel \cite{kaplan2022} keeps all exact constraints satisfied by MCP07 while improving its frequency dependence to better reproduce HEG correlation energies at densities lower than are typical of bulk metals.
Thus rMCP07 and MCP07 differ only for non-zero frequencies.
We assume that the rMCP07 frequency dependence can be further improved, such as by addressing the order-of-limits issue \cite{qian2002}:
\begin{equation}
    \lim_{q\to 0} \left[ \lim_{\omega \to 0} f\mxc(q,\omega) \right] \neq
    \lim_{\omega\to 0} \left[ \lim_{q \to 0} f\mxc(q,\omega) \right].
\end{equation}
Qian and Vignale \cite{qian2002} noted the relevance of the viscosity term in $\mathrm{Im} \, f\mxc(q=0,\omega)$ \cite{conti1999} at high density.
The authors have also demonstrated consequences of the order-of-limits problem in prior work \cite{kaplan2022}.

The screened Coulomb interaction's form is 
\begin{equation}
    W(q,\omega) = \vc(q) \epsilon^{-1}(q,\omega),
\end{equation}
where both sides can have a positive or negative sign.
The potential seen by a test charge is
\begin{equation}
    \delta v_\mathrm{screen}(q) = \delta v\ext(q) + \vc(q) \delta n(q),
\end{equation}
with $\delta v\ext(q)$ the change in the external potential seen by a test electron.
With the change in the electronic density $\delta n(q) = \chi(q) \delta v\ext(q)$, the test-charge--test-charge (TCTC) dielectric function $\epsilon$ is given by
\begin{equation}
    \epsilon^{-1}(q) = 1 + \vc(q) \chi(q).
    \label{eq:DF_TCTC}
\end{equation}
When Eq. (\ref{eq:DF_TCTC}) is used in $W$, it yields that effective interaction which expresses the classical electrostatic potential induced by the density response.

The stability of the HEG \cite{giuliani2005} requires $\chi(q) < 0$ and $\epsilon^{-1}(q) < 1$.
The Kohn-Sham potential created by a test charge is 
\begin{equation}
    \delta v\mks(q) = \delta v\ext(q) + \left[\vc(q) + f\mxc(q) \right] \chi(q) \delta v\ext(q).
\end{equation}
The test-charge--test-electron (TCTE) dielectric function is then given by
\begin{equation}
    \et^{-1}(q) = 1 + \left[\vc(q) + f\mxc(q) \right] \chi(q).
    \label{eq:DF_TCTE}
\end{equation}
It should be noted that the response function $\chi(q)$ entering Eqs. (\ref{eq:DF_TCTC}) and (\ref{eq:DF_TCTE}) is the static limit of the interacting response function in Eq. (\ref{eq:chi_inter}).

\begin{figure}
    \centering
    \includegraphics[width=\columnwidth]{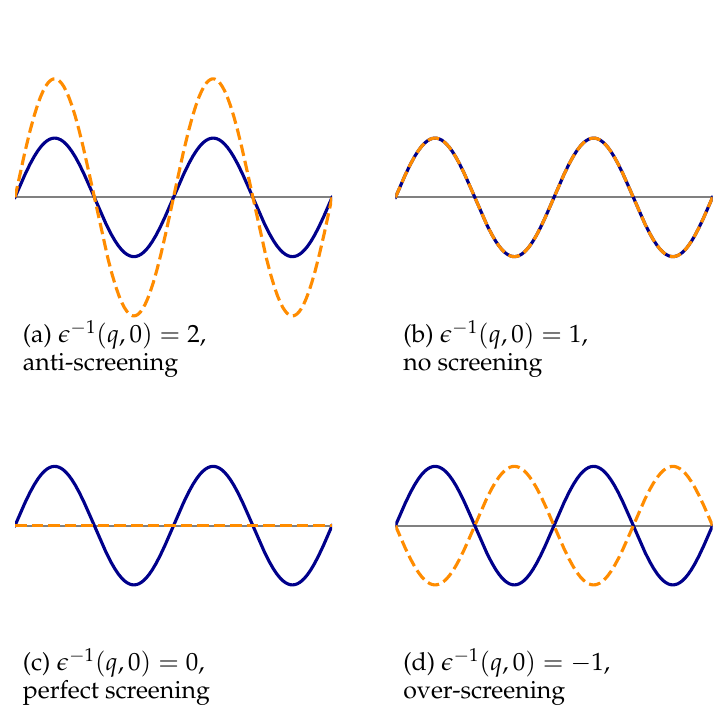}
    \caption{Simple examples of screening in solids.
    In all panels, the external potential is modeled as a sine curve $\delta v\ext(q) = v_0 \sin(q/\kf)$ and plotted as the solid blue curve.
    The screened interaction $W$ is plotted as the dashed orange curve.
    Panel (a) plots anti-screening, whereby electronic correlation \emph{enhances} the amplitude of the perturbation; (b) plots no screening; (c) plots perfect screening, as in a classical metal; and (d) plots over-screening, whereby the screened interaction is $180^0$ out-of-phase with $\delta v\ext$.
    }
    \label{fig:screen_cartoon}
\end{figure}

Different prototypes of screening are illustrated schematically in Fig. \ref{fig:screen_cartoon}.
These examples include anti-screening, where the electronic response enhances $\delta v\ext$; perfect screening whereby $W = 0$, as in a classical metal; and over-screening, where the electronic response is $180^0$ out of phase with $\delta v\ext$.

When $\chi(q) \leq 0$ \emph{and} $ f\mxc(q) \leq 0$, $\et^{-1} \geq \epsilon^{-1}$ by Eqs. (7) and (9). 
At a certain critical density, the TCTC $\epsilon^{-1}$ can become negative \cite{takada2016}, although a negative $\epsilon^{-1}$ does not necessarily indicate an instability in an infinite HEG.
A negative TCTC dielectric function (equivalently, a negative compressibility) makes the screened Coulomb interaction among like test charges attractive at some separations.

\section{Static Screening: Results and Discussion}

\begin{figure*}
    \centering
    \includegraphics[width=6in]{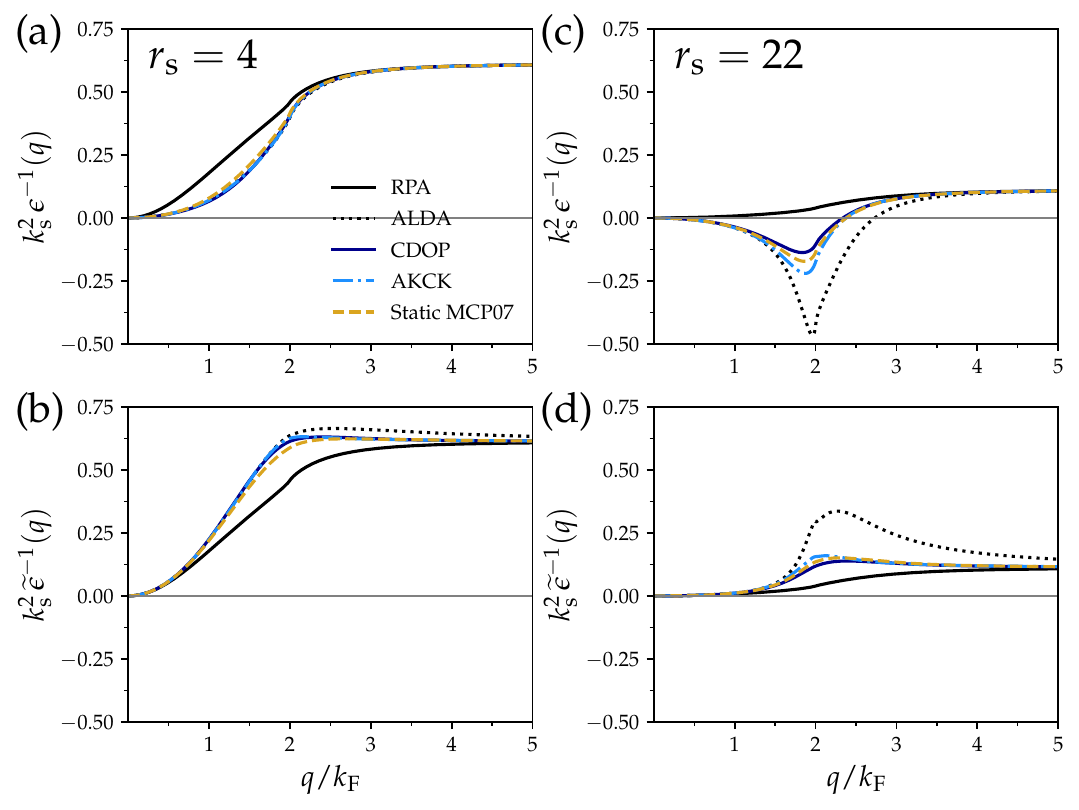}
    \caption{The TCTC $\epsilon^{-1}$ [panels (a) and (c)] and TCTE $\et^{-1}$ [(b) and (d)] dielectric functions, for $\rs = 4$ [panels (a)-(b)] and 22 [(c)-(d)].
    The kernels used are the RPA (solid black), ALDA (dotted gray), Corradini \textit{et al.}\cite{corradini1998} (CDOP, solid dark blue), AKCK\cite{kaplan2023} (dash-dotted light blue), and MCP07\cite{ruzsinszky2020} (dashed yellow).
    The ALDA uses the Perdew-Wang \cite{perdew1992} parameterization of the HEG xc energy density.
    While the TCTE dielectric functions are qualitatively similar at such different densities, the TCTC dielectric functions differ markedly.
    As shown in Eq. (\ref{eq:DF_TCTE_small_q}), $\et^{-1} \approx q^2/\ks^2$, where $\ks = (4\kf/\pi)^{1/2}$ is the Thomas-Fermi screening wave-vector.
    Thus the dielectric functions are scaled by $\ks^2$.
    As $q/\kf \to \infty$, both $\et^{-1}, \, \epsilon^{-1} \to 1$. 
    }
    \label{fig:DF_comp}
\end{figure*}

The dielectric function \cite{singwi1970} is key to capturing correct low-density physics \cite{overhauser1968}.
Therefore, we begin our discussion by comparing $\epsilon^{-1}(q)$ and $\et^{-1}(q)$.
Figure \ref{fig:DF_comp} compares these dielectric functions for two characteristic densities: $\rs = 4$, which is nearly the equilibrium density of the HEG, and typical of the valence electron density of bulk Na; and $\rs = 22$, which Ref. \citenum{takada2016} identified as featuring a ghost exciton.
For $\rs = 4$, both $\epsilon^{-1}(q)$ and $\et^{-1}(q)$ are non-negative.
At $\rs = 22$, the RPA dielectric function (which neglects all xc effects) remains non-negative at all $q$, but inclusion of xc effects yields a sharp dip in $\epsilon^{-1}(q \approx 2\kf)$ towards negative values.
$\kf = (3\pi^2 n)^{1/3}$ is the Fermi wave-vector.
This behavior is consistent across the ALDA, Corradini \textit{et al.}\cite{corradini1998} (CDOP), AKCK \cite{kaplan2023}, and MCP07\cite{ruzsinszky2020} static kernels, although it appears that ALDA largely overestimates xc effects.
Including wave-vector dependence in the AKCK, MCP07, and CDOP kernels reduces the over-screening tendency of ALDA, in that order.
All three kernels obey the same physical constraints, but the more rapid transition to large-$q$ asymptotic behavior in AKCK yields markedly larger over-screening than do MCP07 or CDOP.

Figure \ref{fig:DF_comp} shows that both $\epsilon^{-1}$ and $\et^{-1}$ vanish quadratically as $q \to 0$.
For small-$q$,
\begin{equation}
    \lim_{q \to 0} \chi\mks(q,\omega=0) = -\frac{\kf}{\pi^2}\left[1 - \frac{x^2}{12}
    + \mathcal{O}\left(x^4\right) \right],
\end{equation}
where $x \equiv q/\kf$.
For any reasonable approximation to the HEG kernel,
\begin{equation}
    f\mxc(q) = f\mxc(0) + D\mxc(\rs)q^2 + \mathcal{O}(q^4).
\end{equation}
Thus
\begin{equation}
    \et(q) = \frac{\ks^2}{q^2}\left[ 1 + 
    \left(\frac{\kf^2 f\mxc(0)}{4\pi} - \frac{1}{12} + \frac{\kf^2}{\ks^2} \right)x^2 + \mathcal{O}(x^4)
    \right],
\end{equation}
where $\ks \equiv (4\kf/\pi)^{1/2}$ is the Thomas-Fermi screening length.
Thus one can see that the coefficient of ${q^2}$ in the expansion $\et^{-1}(q)$
\begin{equation}
    \et^{-1}(q) \approx \left(\frac{q}{\ks} \right)^2 
    + \left[\frac{4}{3\pi^2 \ks^2 } - 1 - \frac{\ks^2 f\mxc(0)}{4\pi} \right] \left(\frac{q}{\ks}\right)^4
    \label{eq:DF_TCTE_small_q}
\end{equation}
is always positive.
After much simplification,
\begin{equation}
    \epsilon^{-1}(q) \approx \left[1 + \frac{\ks^2 f\mxc(0)}{4\pi} \right]\left(\frac{q}{\ks}\right)^2,
\end{equation}
which is strictly positive for RPA.
When $f\mxc(0) = f\mxc^\text{ALDA}$, this coefficient changes continuously from positive to negative as $\rs$ passes through the critical $\rs \approx 5.25$ where the HEG total compressibility becomes negative.

The similarity of the peak is supported by the density dependence in CDOP, AKCK, and the static MCP07 kernels.
Furthermore, the dip at $q \approx 2 \kf$ is easily explained by the $2\kf$ ``hump'' phenomenon \cite{overhauser1970,utsumi1980,kaplan2023}.
The AKCK kernel's improved fit to QMC data\cite{kukkonen2021,kaplan2023} suggests greater reliability than CDOP.
The dip is potentially relevant to phonon dispersion and superconductivity in real materials \cite{wang1984,shirron1986}.

\begin{figure}
    \centering
    \includegraphics[width=
    \columnwidth]{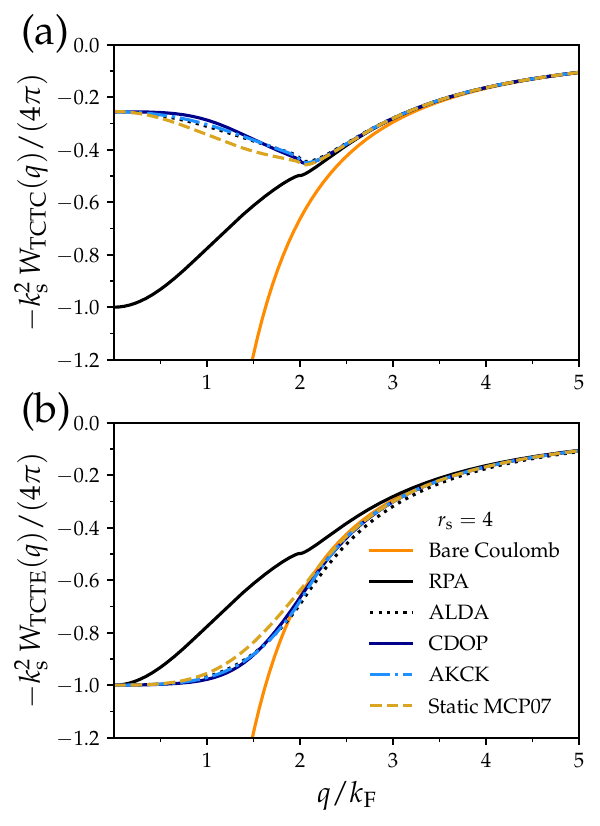}
    \caption{Screened interactions (a) $W_\mathrm{TCTC}(q) = \vc(q)\epsilon^{-1}(q)$ and (b) $W_\mathrm{TCTE}(q) = \vc(q)\et^{-1}(q)$ for a $\rs=4$ HEG.
    The color scheme is the same as used previously, while the bare Coulomb interaction $\vc(q)=4\pi/q^2$ is the solid orange curve.
    Eq. (\ref{eq:DF_TCTE_small_q}) suggests that $W_\mathrm{TCTE}(q) \approx 4\pi/\ks^2$  at small-$q$, thus we scale $W$ by $-\ks^2/(4\pi)$.
    }
    \label{fig:W_q_rs_4}
\end{figure}

$\epsilon^{-1}(q)$ and $\et^{-1}(q)$ are important starting points for our analysis, but a greater understanding is achieved via the effective interaction $W(q)$.
Here, we plot $W$ in both Fourier and real space.
Within TDDFT, $W$ represents how the bare Coulomb interaction is screened.
In BSE, $W$ does not have a similarly simple interpretation.
The negativity of ALDA as shown in Fig. \ref{fig:DF_comp} suggests that the RPA-level $\epsilon^{-1}(q)$ may lead to an improved $W$ within BSE \cite{koskelo2023}.

\begin{figure}
    \centering
    \includegraphics[width=
    \columnwidth]{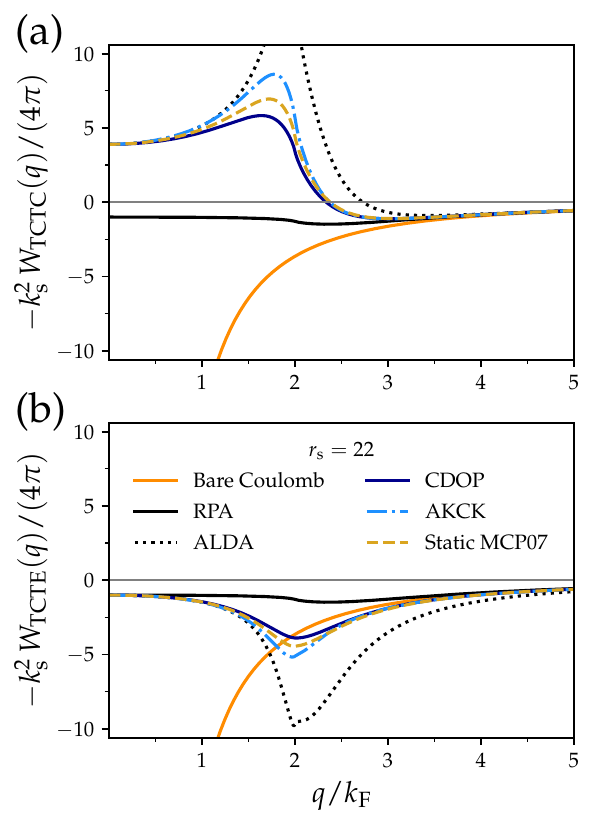}
    \caption{Screened interactions (a) $W_\mathrm{TCTC}(q) = \vc(q)\epsilon^{-1}(q)$ and (b) $W_\mathrm{TCTE}(q) = \vc(q)\et^{-1}(q)$ for a $\rs=22$ HEG.
    }
    \label{fig:W_q_rs_22}
\end{figure}

The shortcomings of both RPA and ALDA $\epsilon^{-1}(q)$, and the necessity of using $\et^{-1}(q)$ in $W$ rather than $\epsilon^{-1}$, were pointed out in Ref. \citenum{koskelo2023}.
Our Figs. \ref{fig:W_q_rs_4}--\ref{fig:W_q_rs_22}, and Appendix Fig. \ref{fig:W_r_rs_4}, confirm this observation while adding more insight with our recent models.
$W_\mathrm{TCTC}$ is the effective interaction at the linear response level between external point charges of charge $+1$ and $-1$, and $W_\mathrm{TCTE}$ is the corresponding effective interaction of Kohn-Sham potential seen by an electron in the presence of an external point charge of charge $+1$.
The most correct results are from MCP07 and AKCK.
Note that only $W_\mathrm{TCTC}$ shows strong over-screening, and then only at low density ($\rs = 22$).
$W_\mathrm{TCTE}$ at $\rs = 22$ exhibits the $q\approx 2\kf$ peak emphasized by Overhauser \cite{overhauser1970}, suggesting an attractive electron-hole interaction.

\begin{figure}
    \centering
    \includegraphics[width=
    \columnwidth]{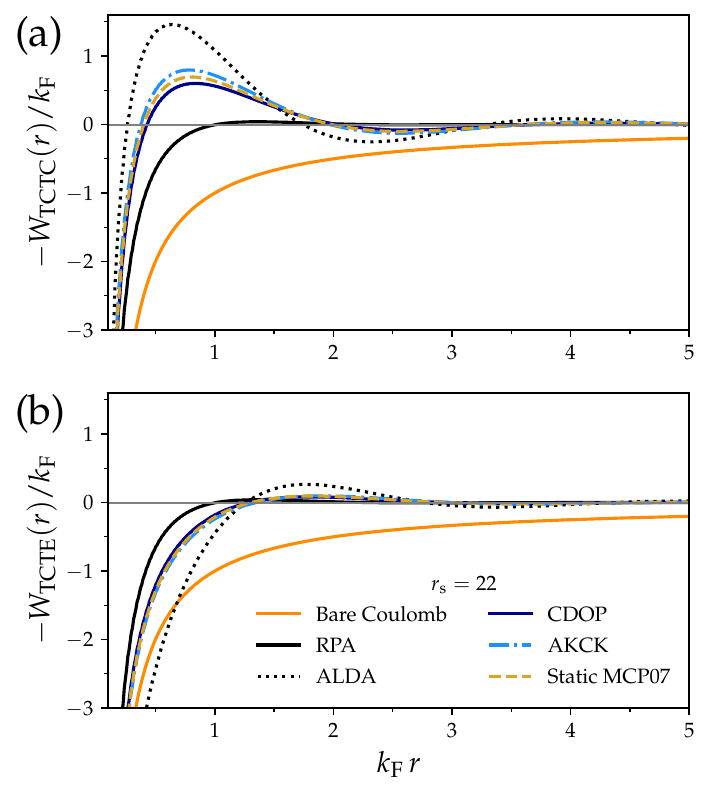}
    \caption{Screened interactions (a) $W_\mathrm{TCTC}(r)$ and (b) $W_\mathrm{TCTE}(r) $ for a $\rs=22$ HEG obtained through a numeric Fourier transform.
    The bare Coulomb interaction $1/r$ is the solid orange curve.
    A real-space analog of Fig. \ref{fig:W_q_rs_4} is given in Appendix Fig. \ref{fig:W_r_rs_4}.
    }
    \label{fig:W_r_rs_22}
\end{figure}

Appendix \ref{app:2DHEG} computes the dielectric function and screened interaction (in both reciprocal and real space) for the 2D HEG using the RPA, the ALDA, and a static kernel (DPGT) \cite{davoudi2001} constructed for the HEG.
For the 2D HEG, the ALDA screened interaction is attractive for $\kf r \leq 1$ at much higher densities, $\rs \approx 4$, compared to the corresponding $\rs$ for the 3D HEG.
These values of $\rs$ for the 2D HEG are experimentally accessible.
More, the static DPGT kernel predicts slightly enhanceed screening over the ALDA.
Further details are provided in Appendix \ref{app:2DHEG}.

\section{Dynamic screening, excitations, and quasi-excitations}

Equations (\ref{eq:chi_inter}) and (\ref{eq:DF_chi}) include frequency dependence.
The frequency-dependent analogs of Eqs. (\ref{eq:DF_TCTC}) and (\ref{eq:DF_TCTE}) are
\begin{align}
    \epsilon^{-1}(q,\omega) &= 1 + \vc(q) \chi(q,\omega), \label{eq:DF_TCTC_dyn} \\
    \et^{-1}(q,\omega) &= 1 + \left[\vc(q) + f\mxc(q,\omega) \right]\chi(q,\omega).
    \label{eq:DF_TCTE_dyn}
\end{align}
At non-zero frequency, the quantities $\chi\mks(q,\omega)$, $\chi(q,\omega)$, $f\mxc(q,\omega)$, and the dielectric functions can be complex.
When $f\mxc$ is set to zero, we obtain the RPA $\epsilon^{-1} = \et^{-1} = \epsilon_\mathrm{RPA}^{-1}$.
From Eqs. (\ref{eq:chi_inter}), (\ref{eq:DF_chi}), (\ref{eq:DF_TCTC_dyn}), and (\ref{eq:DF_TCTE_dyn}), we find
\begin{equation}
    \epsilon^{-1}(q,\omega) = \et^{-1}(q,\omega) 
    \left[1 - f\mxc(q,\omega) \chi\mks(q,\omega) \right].
\end{equation}

Another view of low-density regimes is afforded by plasmon dispersion \cite{giuliani2005,ullrich2014}.
Plasmon dispersion relations are typically plotted until the wave-vector region where plasmons can decay into single-particle excitations \cite{tatarczyk2001}, the particle-hole continuum.
The poles of the interacting response function numerically deliver plasmon dispersion relations \cite{nepal2020}: $\omega_\mathrm{p}(q)$ is found by fixing a real wave-vector $q$ and searching over complex frequencies $\omega$ for the one that zeros-out $\et(q,\omega)$.
MCP07 and rMCP07 are complex, dynamic xc kernels, but an approximate plasmon dispersion can be produced by the zeros of $\et(q,\omega)$ for static kernels.
The particle-hole continuum is bounded by the curves $q^2/2 \pm \kf q$.
MCP07 and rMCP07 produce finite plasmon lifetimes for $q < \kf$, indicating multiple-decay channels outside the particle-hole continuum region.
Ref. \citenum{ruzsinszky2020} found a positive plasmon dispersion at $\rs = 4$ and a negative dispersion at $\rs = 69$, where a plasmon-like density fluctuation drops to zero frequency \cite{perdew2021}.

\begin{figure}
    \centering
    \includegraphics[width=\columnwidth]{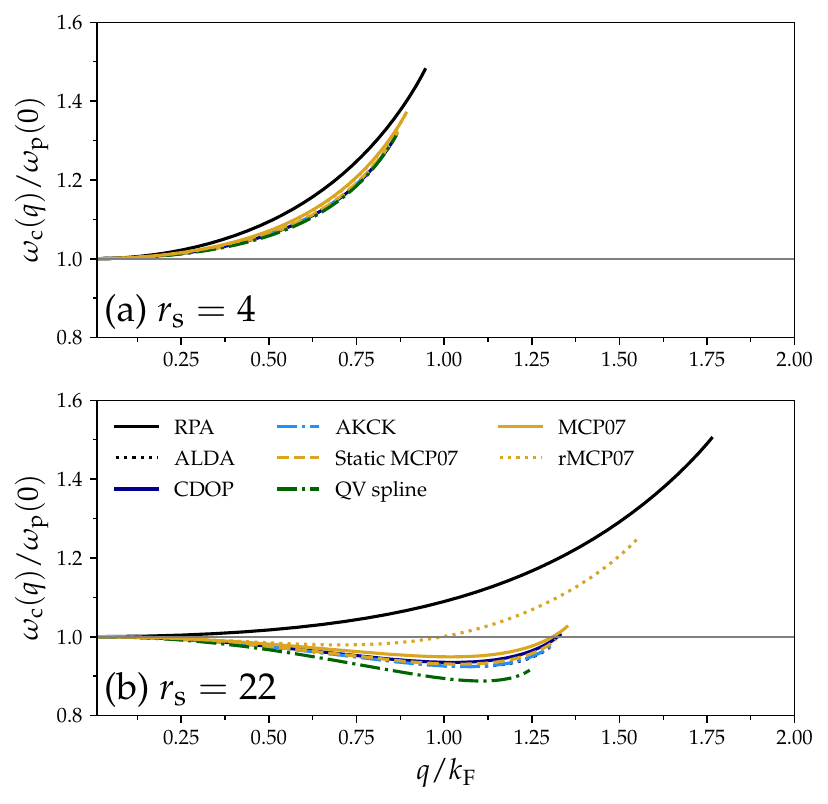}
    \caption{The plasmon-like dispersion $\omega_\mathrm{c}(q)$ (as defined in the text) computed from the RPA (solid black), ALDA (dotted black), CDOP \cite{corradini1998} (solid dark blue), AKCK \cite{kaplan2023} (dash-dotted light blue), static MCP07 (dashed yellow) and dynamic MCP07 (solid yellow), rMCP07 \cite{kaplan2023} (dotted yellow), and Qian-Vignale dynamic LDA \cite{qian2002} (QV, dash-dotted green) TCTE dielectric functions.
    $\omega_\mathrm{c}$ is scaled by the bulk plasmon frequency $\omega_\mathrm{p}(0) = (4\pi n)^{1/2}$.
    Positive dispersion for $\rs = 4$ is shown in panel (a), and negative dispersion for $\rs = 22$ in panel (b).
    The curves are plotted only until they enter the upper bound of the particle-hole continuum region, bounded from above by $q^2/2 + q \kf$.
    }
    \label{fig:wcq_TCTE}
\end{figure}

The authors of Ref. \cite{koskelo2023} have defined a ``collective mode at real frequency $\omega_\mathrm{c}(q)$'' by solving a somewhat different problem: finding the real frequency that makes $\mathrm{Re} \, \et(q,\omega) = 0$ (again for real-valued $q$).
This delivers the true plasmon dispersion only at very small $q$.
Our Fig. \ref{fig:wcq_TCTE} shows this plasmon-like dispersion, indicating positive ``dispersion'' for $\rs = 4$, and negative ``dispersion'' for $\rs = 22$, as expected.
The curves are cut off when they enter the particle-hole continuum region.

In Fig. \ref{fig:wcq_TCTE}, all model kernels make the $q \to 0$ limit exact, but the xc effects appear strongly in the model kernels and ALDA.
This is only observable for larger $q/\kf$ where the dispersion curves meet the particle-hole continuum cutoff.
However nearly all models predict qualitatively similar shapes for $\omega_\mathrm{c}(q)$.
This is especially relevant for the dynamic LDA of Qian and Vignale (QV) \cite{qian2002}, which does not reduce to the ALDA, and which attempts to approximate two-particle excitations.

\begin{figure}
    \centering
    \includegraphics[width=\columnwidth]{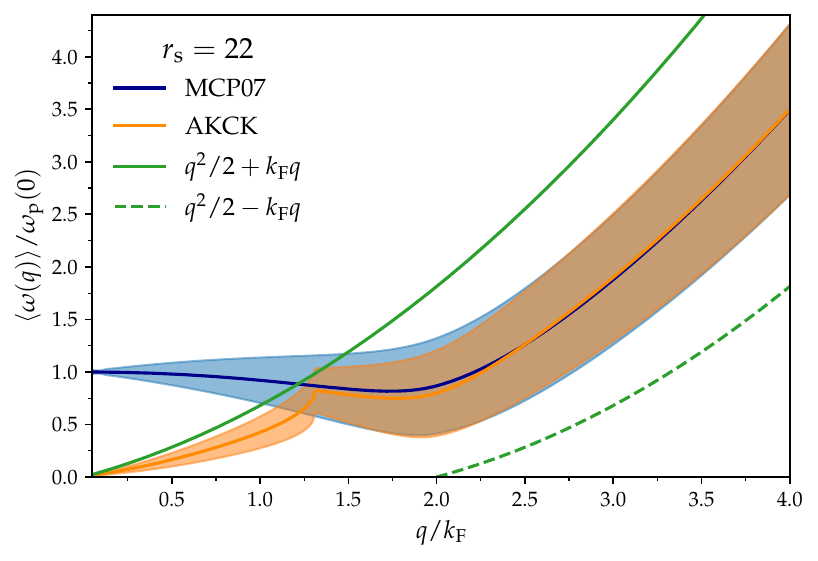}
    \caption{The average frequency of density fluctuation, $\langle \omega \rangle$ of Eq. (\ref{eq:wavg}) (solid blue curve for MCP07 and orange for AKCK), and $\langle \omega(q) \rangle \pm \Delta \omega(q)$ (light blue or orange shaded regions) for an $\rs = 22$ HEG.
    The standard deviation of a density fluctuation, $\Delta \omega(q)$, is defined in Eq. (\ref{eq:wstd}).
    The boundaries of the particle-hole continuum, $\omega_\mathrm{PHC}^{(\pm)} = q^2/2 \pm q \kf$, are plotted as the green curves (solid for $\omega_\mathrm{PHC}^{(+)})$ and dashed for $\omega_\mathrm{PHC}^{(-)})$)}.
    For analogous plots at $\rs = 4, \, 69$, see Appendix Figs. \ref{fig:stat_plas_rs_4} and \ref{fig:stat_plas_rs_69}.
    \label{fig:stat_plas_rs_22}
\end{figure}

The final point of our analysis regards excitations in the low-density regime.
To do so, we use the spectral function $S(q,\omega) = \mathrm{Im} \, \chi(q,\omega)/(\pi n)$ \cite{giuliani2005}.
We follow the definitions of the average frequency of a density fluctuation,
\begin{equation}
    \langle \omega(q) \rangle = \left[ \int_0 ^\infty S(q,\omega)d\omega \right]^{-1} \left[ \int_0 ^\infty \omega S(q,\omega) d\omega\right],
    \label{eq:wavg}
\end{equation}
and variance of a density fluctuation,
\begin{equation}
    \langle \omega^2(q) \rangle = \left[ \int_0 ^\infty S(q,\omega) d\omega \right]^{-1} \left[ \int_0 ^\infty \omega^2 S(q,\omega) d\omega \right],
\end{equation}
presented in Ref. \citenum{perdew2021}.
The standard deviation of a density fluctuation is then
\begin{equation}
    \Delta \omega(q) = \left[ \langle \omega^2(q) \rangle - \langle \omega(q) \rangle^2 \right]^{1/2}.
    \label{eq:wstd}
\end{equation}
Figure \ref{fig:stat_plas_rs_22} (for $\rs = 22$) and Appendix Figs. \ref{fig:stat_plas_rs_4} and \ref{fig:stat_plas_rs_69} (for $\rs = 4$ and 69, respectively), present these quantities for a wide range of $q$.
At $\rs = 69$ (Appendix Fig. \ref{fig:stat_plas_rs_69}), a charge-density wave is observed \cite{perdew2021} as $\langle\omega\rangle$ dropping towards zero frequency near $q = 2 \kf$.

In Fig. \ref{fig:stat_plas_rs_22}, the solid blue curve is $\langle \omega(q) \rangle$ from MCP07, and the light-blue shaded region is bounded by $\langle \omega(q) \rangle \pm \Delta \omega(q)$.
The orange curve and shaded regions are identical quantities computed with AKCK.
The green curves are the bounds of the particle-hole continuum (PHC), $\omega_\mathrm{PHC}^{(\pm)} q^2/2 \pm \kf q$.
The negative plasmon dispersion is clearly evident at $\rs = 22$ with MCP07, and at $\rs = 69$ for AKCK and MCP07.
$\Delta \omega$ is an indicator of collective excitations, as we discuss below.
While it is not surprising that, for $\langle\omega(q)\rangle > q^2/2 + q \kf$, $\Delta \omega < \langle\omega(q)\rangle$, it is surprising that this inequality holds even when $\langle\omega(q)\rangle $ enters the particle-hole continuum.
We conclude that any density fluctuation within the particle-hole continuum gains some collective nature, and resembles plasmon excitations of the bound electron-hole pairs.
This is not a true excitation [i.e., a pole of $\chi(q,\omega)$], so we call it a quasi-excitation.

The lifetime of a collective excitation can be estimated by the full-width at half-maximum (FWHM) of the spectral function \cite{lewis2019}.
Appendix Fig. \ref{fig:spectral_mcp07_rs_69} plots the MCP07 spectral function at $\rs = 69$, showing the positions of $\omega_\mathrm{PHC}^{(\pm)}$ and the FWHM.
For $q \lesssim 1.5 \kf$, $\omega_\mathrm{PHC}^{(+)}$ lies below the FWHM frequencies.
This is entirely consistent with Appendix Fig. \ref{fig:stat_plas_rs_69}, where $\langle \omega (q) \rangle$ and $\omega_\mathrm{PHC}^{(+)}$ cross, and indicates a plasmon of finite lifetime.
For $q \approx 2\kf$, the spectral function diverges for $\omega \to 0$, and no FWHM can be identified.
This indicates an excitation of extremely long lifetime, as suggested by $\langle \omega \rangle$ tending to zero there.
For $q \gtrsim 2.5\kf$, a FWHM can again be identified, but the excitations again lie within the PHC, consistent with $\langle \omega(q) \rangle - \langle \Delta \omega(q) \rangle$ crossing $\omega_\mathrm{PHC}^{(-)}$ there.
The approximate MCP07 spectral function FWHM and $\langle \Delta \omega \rangle$ are plotted for $\rs = 22$ and 69 in Appendix Fig. \ref{fig:lifetime}; the two appear to be generally uncorrelated.

However, when the spectral function, at fixed $q$, can be globally approximated as a Gaussian, both $\langle \omega(q) \rangle$ and $\langle \Delta \omega(q) \rangle$ can be analytically related to the functional form of the Gaussian, as shown in Appendix \ref{app:gaussian_stats}.
Such expressions allow one to trivially relate the plasmon lifetime (FWHM) to $\langle \Delta \omega \rangle$.
Appendix \ref{app:gaussian_stats} shows that the plasmon lifetime varies between roughly $(2.355)\langle \Delta \omega \rangle$ when $S(q,\omega)$ is almost constant in $\omega$, and $(3.906)\langle \Delta \omega \rangle$ when $S(q,\omega)$ is very sharply peaked.

Ghost excitations are thus also quasi-excitations.
They are poles of the \emph{screened} response function $\chi_\mathrm{scr}(q,\omega)$ \cite{takayanagi1997} that gives the linear response of the density to the \emph{screened} interaction (the sum of the external perturbing field and the Hartree potential it induces).
Since
\begin{equation}
    \chi(q,\omega) = \chi_\mathrm{scr}(q,\omega) \left[1 - \vc(q) \chi_\mathrm{scr}(q,\omega)  \right]^{-1},
\end{equation}
a pole of $\chi_\mathrm{scr}$ is not necessarily a pole of $\chi$.

\section{Conclusions}

We have extensively explored the physics of low densities with beyond-RPA approximations in linear response TDDFT.
Exchange-correlation (xc) effects play a relevant role in the low-density regime, as they are exemplified through the screened Coulomb interaction, plasmon dispersion, and quasi-excitations.
Real metallic materials can be modeled by the HEG, as we do in this paper.
Apart from the simplest ALDA kernel, we have used the CDOP \cite{corradini1998}, AKCK \cite{kaplan2023}, MCP07 \cite{ruzsinszky2020}, rMCP07 \cite{kaplan2022}, and QV \cite{qian2002} kernels, so that both static and dynamic kernels are represented.
All code used in this work is publicly accessible \cite{heg_screen_repo}.

The effective interaction at low densities displays negative screening from our static xc kernels, as a consequence of the small-$q$ expansion of $f\mxc$.
The effective interaction $W_\mathrm{TCTE}$ in real space shows attraction between particles and holes below $r = 1/\kf$ for all xc models, confirming the possible existence of excitons in metals at short range.

The densities needed to observe excitons in 3D metals, $\rs \gtrsim 20$, are likely experimentally inaccessible.
However, Appendix \ref{app:2DHEG} demonstrates that the screened interaction becomes attractive at experimentally accessible densities, $\rs \gtrsim 4$ in the \emph{2D} HEG.

TCTE plasmon-like dispersion undoubtedly reveals the discrepancy between RPA and beyond-RPA approximations.
Model xc kernels make the plasmon dispersion negative even for small $q/\kf$, while xc effects strengthen near the particle-hole continuum.

We also performed an analysis regarding the nature of excitations in the plasmon and particle-hole continuum regions.
The spectral function was used to plot the average frequency and variance in the frequency of a density fluctuation at wave-vector $q$.
These latter quantities are constituents of the standard deviation of the average frequency $\Delta \omega$ that we used to study quasi-excitations in the particle-hole continuum.
From $\Delta \omega$, we find that the density fluctuation within the particle-hole continuum gains some collective nature and resembles collective plasmon excitations of the bound electron-hole pairs.

Our model kernels indicate that the physics of the low-density regime is rich.
Moreover, these physically-constrained xc kernels can be useful tools to study emergent phenomena in the low-density regime of real materials \cite{mott1968,schleife2011,ohtomo2004,li2007}.

\begin{acknowledgments}
The work of AR was supported by the U.S. Department of Energy (DOE), Office of Science, Office of Basic Energy Sciences (BES), under Award Number DE-SC0021263, and by the start-up funding from Tulane University. 
ADK acknowledges support by the DOE-BES Materials Project Program, Contract No. KC23MP.
Helpful discussions with Prof. John P. Perdew are largely appreciated.
\end{acknowledgments}

\section*{Data Availability}
The data that support the findings of this study are openly available at the code and data repository\cite{heg_screen_repo}.

\appendix
\section{Additional Figures}

\begin{figure}[h]
    \centering
    \includegraphics[width=
    \columnwidth]{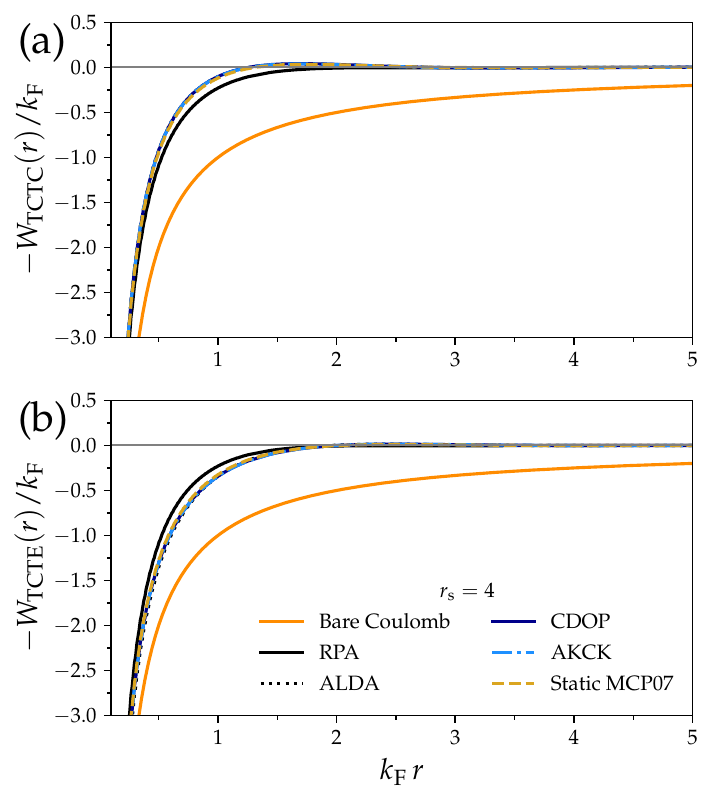}
    \caption{Screened interactions (a) $W_\mathrm{TCTC}(r)$ and (b) $W_\mathrm{TCTE}(r) $ for a $\rs=4$ HEG obtained through a numeric Fourier transform.
    The bare Coulomb interaction $1/r$ is the solid orange curve.
    }
    \label{fig:W_r_rs_4}
\end{figure}

\begin{figure}
    \centering
    \includegraphics[width=\columnwidth]{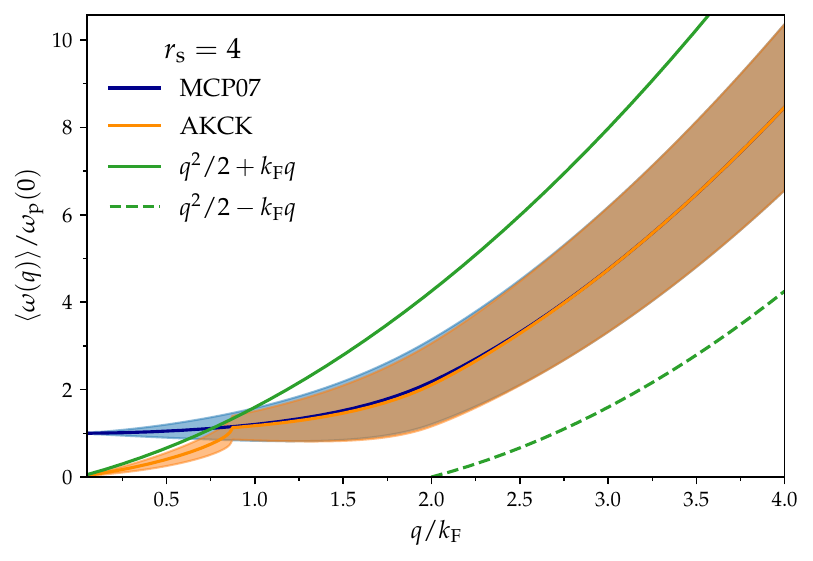}
    \caption{The average frequency of density fluctuation, $\langle \omega \rangle$ of Eq. (\ref{eq:wavg}) (solid blue curve for MCP07 and orange for AKCK), and $\langle \omega(q) \rangle \pm \Delta \omega(q)$ (light blue shaded region) for an $\rs=4$ HEG.
    The standard deviation of a density fluctuation, $\Delta \omega(q)$, is defined in Eq. (\ref{eq:wstd}).
    The boundaries of the particle-hole continuum, $\omega_\mathrm{PHC}^{(\pm)}) = q^2/2 \pm q \kf$, are plotted as the green curves (solid for $\omega_\mathrm{PHC}^{(+)})$ and dashed for $\omega_\mathrm{PHC}^{(-)})$).
    }
    \label{fig:stat_plas_rs_4}
\end{figure}

\begin{figure}
    \centering
    \includegraphics[width=\columnwidth]{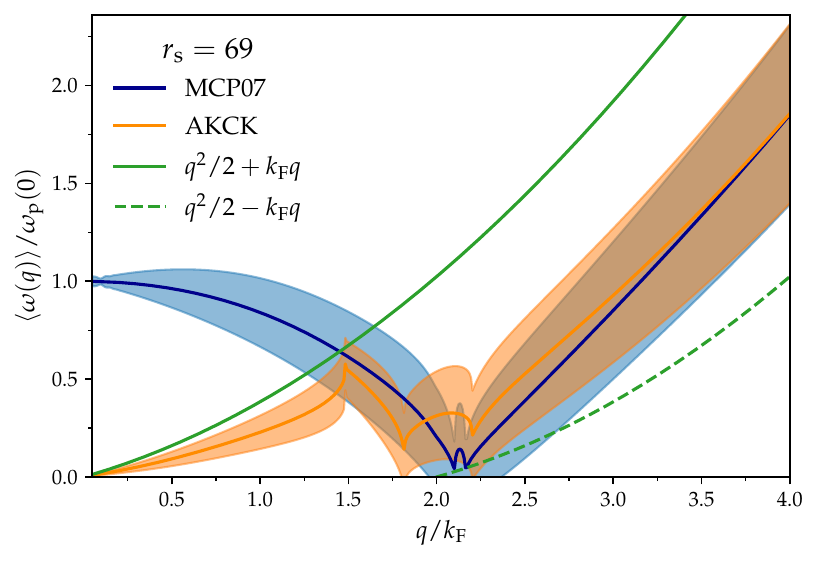}
    \caption{Analogous to Fig. \ref{fig:stat_plas_rs_4}, but for an $\rs = 69$ HEG.
    }
    \label{fig:stat_plas_rs_69}
\end{figure}

\begin{figure}
    \centering
    \includegraphics[width=\columnwidth]{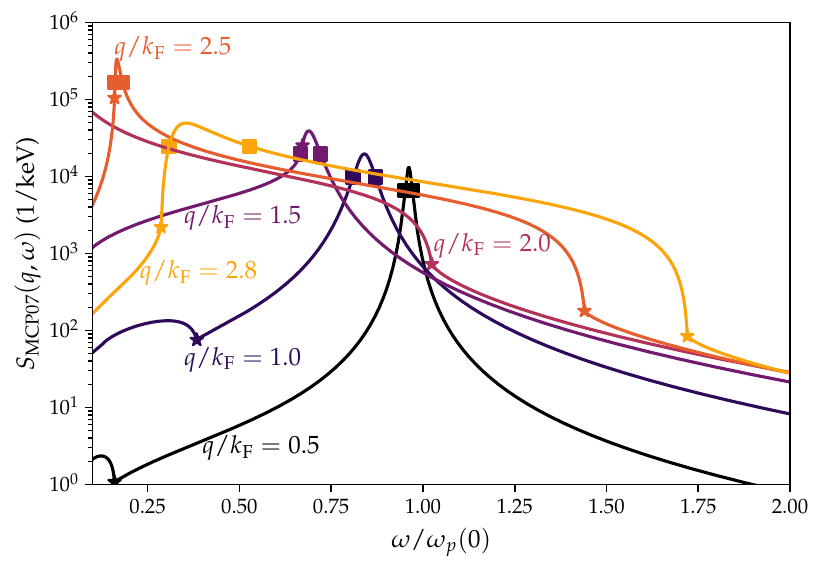}
    \caption{The MCP07 spectral function, or dynamic structure factor $S(q,\omega)=-\mathrm{Im}\, \chi(q,\omega)/(\pi n)$ for a few values of $q$.
    The stars indicate the frequencies $\omega_\mathrm{PHC}^{(\pm)} = q^2/2 \pm q \kf$ of particle-hole continuum (PHC) excitations, and the squares indicate the approximate position of the half-width of the spectral function at half maximum (HWHM).
    The full-width at half maximum (FWHM) of the spectral function can be used to estimate the lifetime of a collective excitation.
    }
    \label{fig:spectral_mcp07_rs_69}
\end{figure}

\begin{figure}
    \centering
    \includegraphics[width=\columnwidth]{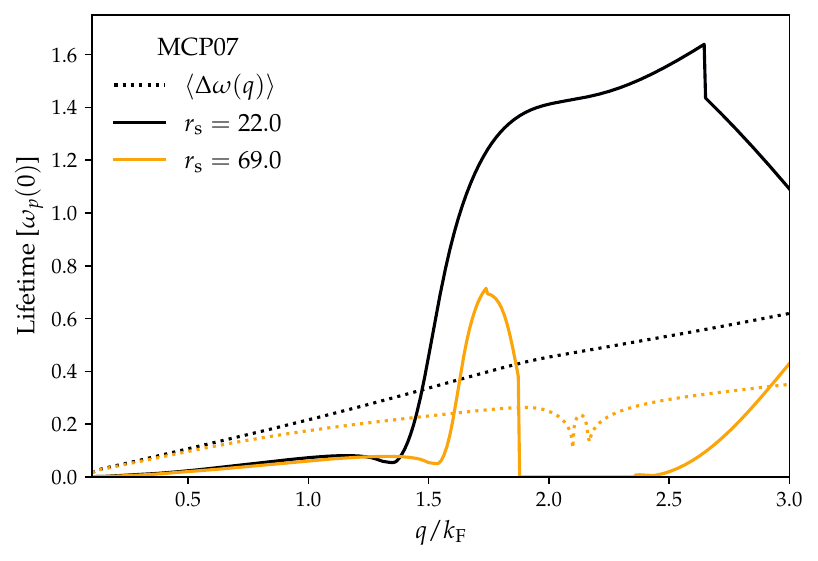}
    \caption{Excitation lifetime, in units of the bulk plasmon frequency $\omega_p(0)= (4\pi n)^{1/2}$, for $\rs = 22$ (yellow) and 69 (black) using MCP07.
    The solid lines compute the full width at half-maximum of the spectral function $S(q,\omega)$ when a maximum can be identified.
    For $1.8 \lesssim q/\kf \lesssim 2.5$, the $\rs=69$ spectral function diverges as $\omega \to 0$, yielding a static charge density wave, as discussed in the text.
    The FWHM is computed via spline fit of $S(q,\omega)$ at fixed $q$; no fit to a lineshape is used.
    The dotted lines estimate the lifetime from the statistical $\langle \Delta \omega(q) \rangle$.
    }
    \label{fig:lifetime}
\end{figure}

Figure \ref{fig:W_r_rs_4} is the real-space analog of Fig. \ref{fig:W_q_rs_4}, and thus plots $W(r)$ for an $\rs = 4$ HEG.
Figures \ref{fig:stat_plas_rs_4} and \ref{fig:stat_plas_rs_69} plot the average frequency and standard deviation of a density fluctuation in $\rs = 4$ and 69 HEGs, analogous to Fig. \ref{fig:stat_plas_rs_22}.
Figure \ref{fig:spectral_mcp07_rs_69} plots the MCP07 spectral function at $\rs = 69$ as a function of real frequency, indicating the positions of the particle-hole continuum frequencies, and full width at half maximum.
Approximate plasmon lifetimes estimated from the spectral function full width at half maximum and from $\langle \Delta \omega \rangle$ are plotted in Fig. \ref{fig:lifetime}.

\pagebreak

\section{Two dimensional electron gas \label{app:2DHEG}}

In $d$ dimensions, the adiabatic local density approximation (ALDA) for the exchange-correlation kernel takes the form
\begin{equation}
    f\mxc(\rs) = \frac{u_d}{d^2} \rs^{d+1} \left[
    \rs \frac{\partial^2 \varepsilon\mxc}{\partial \rs^2} 
    - (d-1) \frac{\partial \varepsilon\mxc}{\partial \rs},
    \right]
\end{equation}
where $\varepsilon\mxc(\rs)$ is the exchange-correlation energy per electron in a $d$-dimensional HEG.
$u_d$ is a constant which gives the volume $\mathcal{V}_d$ of a ball (solid sphere) in $d$ dimensions,
\begin{equation}
    \mathcal{V}_d(R) = u_d R^d,
\end{equation}
thus $u_2 = \pi$ and $u_3 = 4\pi/3$.
In two dimensions, 
\begin{equation}
    \varepsilon\mx(\rs) = -\frac{2^{5/2}}{3\pi \rs}.
\end{equation}
To express the correlation energy per electron, we use the parameterization of Ref. \citenum{attaccalite2002}, which is fitted to fixed-node diffusion Monte-Carlo simulations,
\begin{align}
    \varepsilon\mc(\rs,\zeta=0) &= A + \left(B \rs + C \rs^2 + D \rs^3 \right) \\
    & \times \ln \left[1 + \frac{1}{E \rs + F \rs^{3/2} + G \rs^2 + H \rs^3} \right]
    \nonumber,
\end{align}
where $\zeta$ is the relative spin polarization, $A =-0.1925$, $B = 0.0863136$, $C=0.0572384$, $D=-A\, 
H$, $E = 1.0022$, $F = -0.02069$, $G = 0.33997$, and $H = 0.01747$.
The Lindhard function in two dimensions is \cite{stern1967}
\begin{align}
    \mathrm{Re} \, \chi_0(q,\omega) &= \frac{1}{2\pi z} \left[\phi(z - u) + \phi(z + u) - 2z \right], \\
    \mathrm{Im} \, \chi_0(q,\omega) &= \frac{1}{2\pi z}\left[\psi(z + u) - \psi(z - u)\right], \\
    \phi(x) &= \mathrm{sign}(x) \Theta(|x| - 1) (|x|^2 - 1)^{1/2}, \\
    \psi(x) &= \Theta(1 - |x|) (1 - |x|^2 )^{1/2}, \\
    z &= q/(2\kf), \\
    u &= \omega/(q \kf).
\end{align}

To compare the screened electronic response, we also include the static kernel due to Davoudi, Polini, Giuliani, and Tosi (DPGT) \cite{davoudi2001}.
The DPGT kernel is posed in much the same way as the AKCK kernel, as a smooth interpolation between known asymptotics of the 2D HEG exchange-correlation kernel \cite{davoudi2001},
\begin{align}
    \lim_{q \to 0} f\mxc(\rs, q,\omega=0) &= f\mxc^\mathrm{ALDA}(\rs), \\
    \lim_{q/\kf \gg 1} f\mxc(\rs, q,\omega=0) &= -\frac{2\pi}{\kf} \left[ C_+^\mathrm{(2D)} + B_+^\mathrm{(2D)} \frac{\kf}{q} \right], \\
    C_+^\mathrm{(2D)} &= - \frac{\rs}{2^{1/2}} \left[\varepsilon\mc + \rs \frac{\partial \varepsilon \mc}{\partial \rs} \right], \\
    B_+^\mathrm{(2D)} &+ 1 - g^\mathrm{(2D)}(0).
\end{align}
The 2D HEG on-top pair distribution function is reasonably well-approximated by \cite{polini2001}
\begin{equation}
    g^\mathrm{(2D)}(0) = \frac{1}{2\left[1 + (1.372)\rs + (0.0830) \rs^2 \right]}.
\end{equation}

\begin{figure}
    \centering
    \includegraphics[width=\columnwidth]{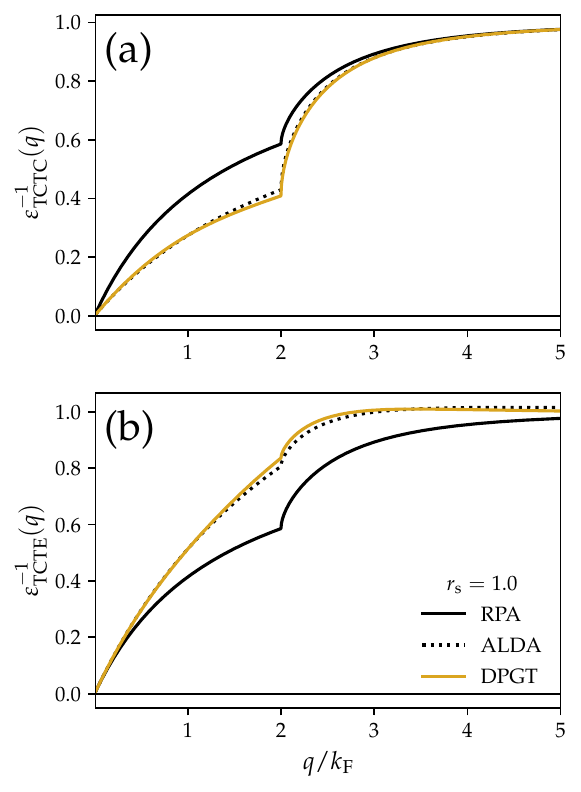}
    \caption{(a) TCTC and (b) TCTE inverse dielectric functions for an $\rs=1$ 2D HEG using the RPA (black, solid), ALDA in the parameterization of Ref. \cite{attaccalite2002} (black, dotted), and the static DPGT kernel \cite{davoudi2001} (yellow, solid).}
    \label{fig:IDF_2D_rs_1}
\end{figure}

\begin{figure}
    \centering
    \includegraphics[width=\columnwidth]{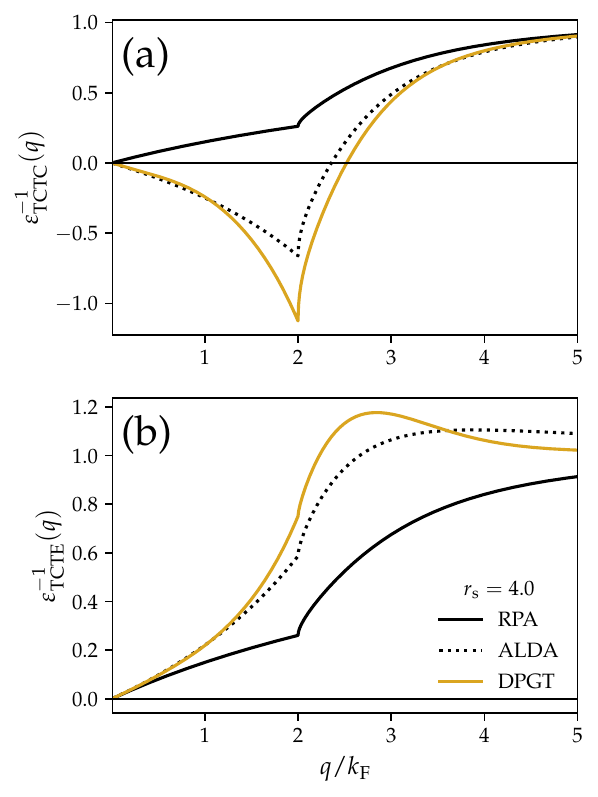}
    \caption{Same as Fig. \ref{fig:IDF_2D_rs_1} but for an $\rs=4$ 2D HEG.}
    \label{fig:IDF_2D_rs_4}
\end{figure}

\begin{figure}
    \centering
    \includegraphics[width=\columnwidth]{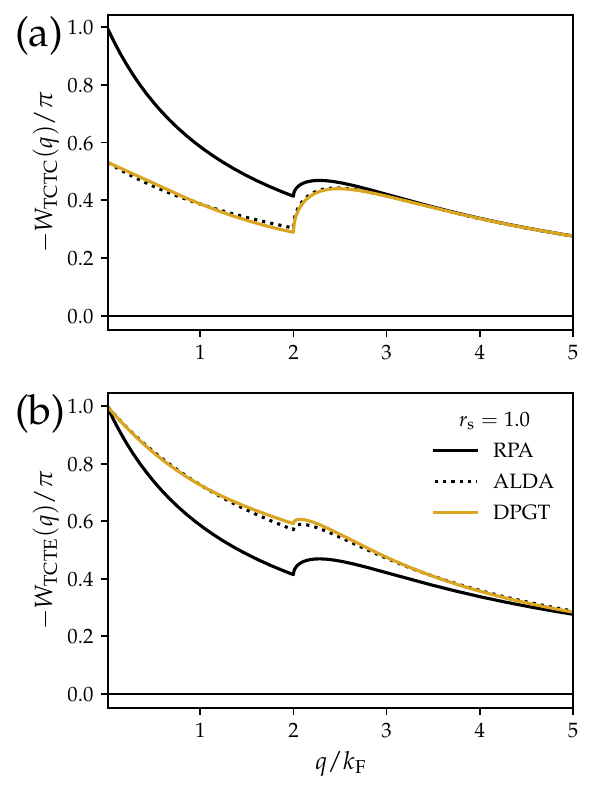}
    \caption{(a) TCTC and (b) TCTE screened interactions $W/\pi$ for an $\rs=1$ 2D HEG using the RPA (black, solid), ALDA in the parameterization of Ref. \cite{attaccalite2002} (black, dotted), and the static DPGT kernel \cite{davoudi2001} (yellow, solid).}
    \label{fig:W_2D_rs_1}
\end{figure}

\begin{figure}
    \centering
    \includegraphics[width=\columnwidth]{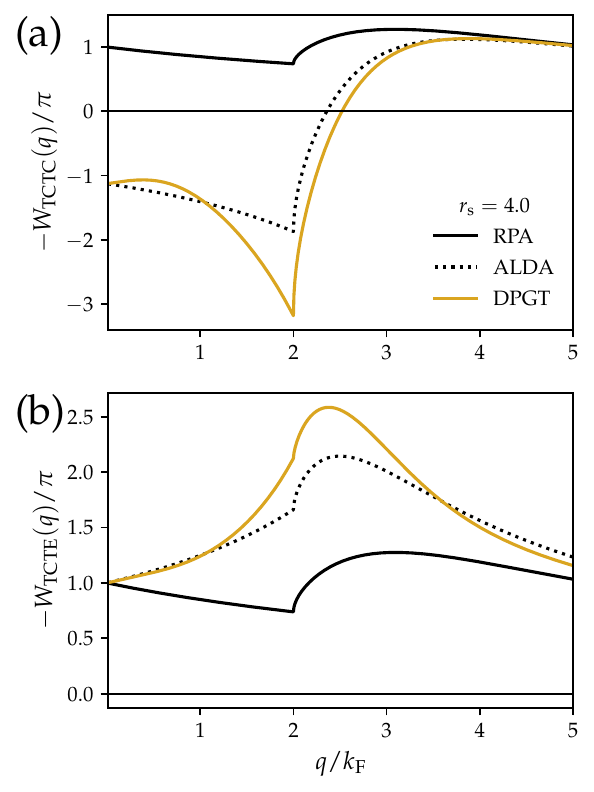}
    \caption{Same as Fig. \ref{fig:W_2D_rs_1} but for an $\rs=4$ 2D HEG.}
    \label{fig:W_2D_rs_4}
\end{figure}

Figure \ref{fig:IDF_2D_rs_1} shows the TCTC and TCTE inverse dielectric functions for the 2D HEG at $\rs = 1$.
Unlike the 3D case, the TCTC dielectric function becomes negative at much smaller $\rs$, as seen in Fig. \ref{fig:IDF_2D_rs_4} for $\rs = 4$.
Also unlike the 3D case, the static DPGT kernel predicts slightly enhanced screening over the ALDA, however both are comparable.
Similar trends are observed for the screened interactions plotted in Figs. \ref{fig:W_2D_rs_1} and \ref{fig:W_2D_rs_4} for $\rs = 1$ and 4 2D HEGs, respectively.
Rigorously, the RPA TCTC screened interaction and all TCTE screened interactions tend to $-\pi$ as $q \to 0$.

\begin{figure}
    \centering
    \includegraphics[width=\columnwidth]{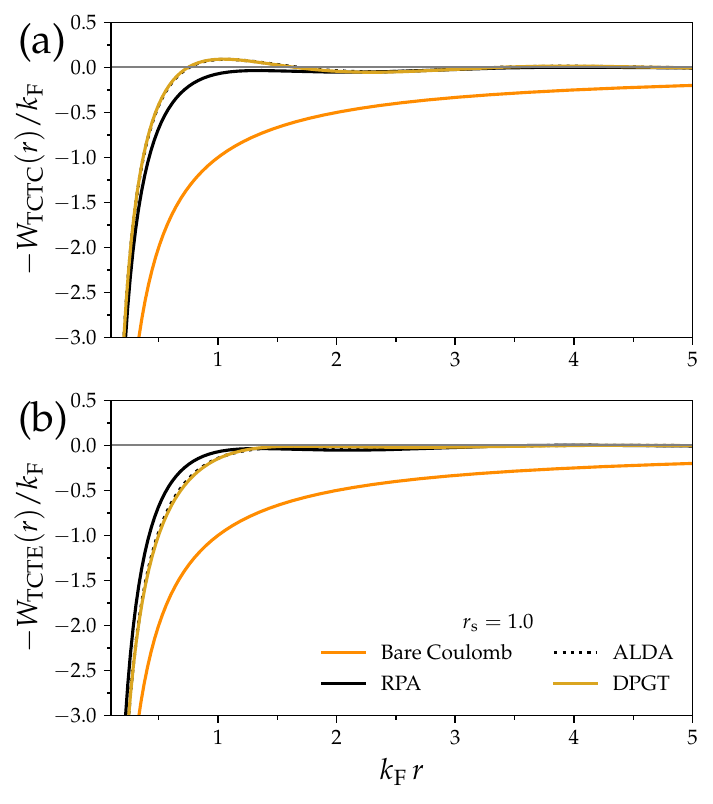}
    \caption{Real-space screened interaction $W(r)$ for an $\rs = 1$ 2D HEG, using the RPA (black, solid), ALDA \cite{attaccalite2002} (black, dotted), and the static DPGT kernel \cite{davoudi2001} (yellow, solid).
    The real-space interaction is obtained by fast Hankel transform.
    }
    \label{fig:W_r_2D_rs_1}
\end{figure}

\begin{figure}
    \centering
    \includegraphics[width=\columnwidth]{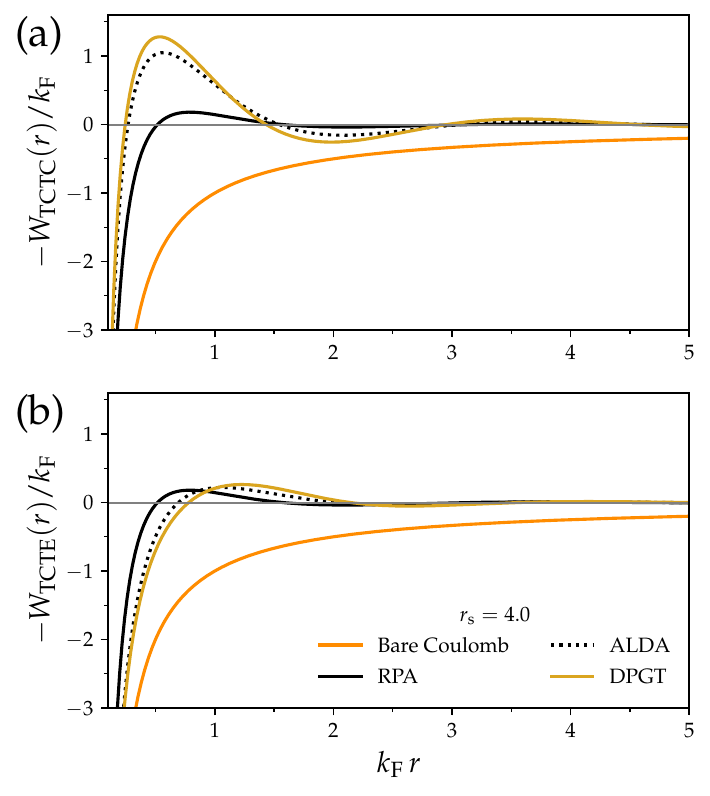}
    \caption{Same as Fig. \ref{fig:W_r_2D_rs_1}, but for an $\rs=4$ 2D HEG.
    }
    \label{fig:W_r_2D_rs_4}
\end{figure}

Last, we plot the real-space screened interactions $W(r)$ for an $\rs = 1$ and 4 2D HEG in Figs. \ref{fig:W_r_2D_rs_1} and \ref{fig:W_r_2D_rs_4}, respectively. 
The real-space interaction is obtained by fast Hankel transform.
Even at $\rs = 1$, the screened interaction from ALDA is weakly attractive, and even more so at $\rs = 4$.
 
\newpage

\section{Physical interpretations of the statistical variables \label{app:gaussian_stats}}

For a fixed value of $q$, suppose the spectral function can be approximated as a Gaussian:
\begin{equation}
    S(q,\omega) \approx S_0(q)\exp\left[- \left(\frac{\omega - \omega_0}{\Delta \omega} \right)^2 \right]
\end{equation}
We choose a Gaussian as a Lorentzian has undefined moments.
This approximation is relevant for estimating the lifetime of an excitation near a maximum $S_0(q)$ of the spectral function, which occurs at frequency $\omega_0$.
The full width at half maximum (FWHM) of the approximate spectral function, which gives the lifetime of an excitation, is
\begin{equation}
    \mathrm{FWHM} = 2 (\Delta \omega) (\ln 2)^{1/2}.
\end{equation}
Our aim now is to make contact with the statistical quantities of Eqs. (\ref{eq:wavg}--\ref{eq:wstd}), which use the following frequency moments,
\begin{equation}
    M_p = \int_0^\infty \omega^p S(q,\omega) d\omega.
\end{equation}
One can show that
\begin{align}
    M_0 &= \frac{\sqrt{\pi}S_0(q) \Delta \omega }{2}\left[1 + \mathrm{erf}\left(\frac{\omega_0}{\Delta \omega} \right) \right], \\
    \frac{M_1}{M_0} &=
        \omega_0
        + \Delta \omega \frac{\exp\left[-\left(\omega_0/\Delta \omega \right)^2 \right]}{\sqrt{\pi} \left[1 + \mathrm{erf}\left(\omega_0 / \Delta \omega \right) \right]}, 
        \label{eq:avg_freq_gauss} \\
    \frac{M_2}{M_0} &= \omega_0^2 + \frac{(\Delta \omega)^2}{2}
        + \omega_0 \Delta \omega\frac{\exp\left[-\left(\omega_0/\Delta \omega \right)^2 \right]}{\sqrt{\pi} \left[1 + \mathrm{erf}\left(\omega_0 / \Delta \omega \right) \right]}.
\end{align}
Equation (\ref{eq:avg_freq_gauss}) is equivalent to $\langle \omega(q) \rangle$ of Eq. (\ref{eq:wavg}), the average frequency of a density fluctuation.
The standard deviation in a density fluctuation of Eq. (\ref{eq:wstd}) is then
\begin{align}
    \langle \Delta \omega(q) \rangle &= \left[ \frac{M_2}{M_0} - \left( \frac{M_1}{M_0}\right)^2\right]^{1/2}, \\
    &= \Delta \omega \left\{ 
        \frac{1}{2} - \frac{\exp\left[-2\left(\omega_0/\Delta \omega \right)^2 \right]}{\pi \left[1 + \mathrm{erf}\left(\frac{\omega_0}{\Delta \omega} \right) \right]^2}  \right. \nonumber \\
         & \left. - \frac{\omega_0}{\Delta \omega} \frac{\exp\left[-\left(\omega_0/\Delta \omega \right)^2 \right]}{\sqrt{\pi} \left[1 + \mathrm{erf}\left(\omega_0 / \Delta \omega \right) \right]}
    \right\}^{1/2}.
\end{align}
Thus we can relate the plasmon lifetime (spectral function FWHM) to the standard deviation in a density fluctuation for an approximately Gaussian spectral function as
\begin{align}
    \mathrm{FWHM} =& 2(\ln 2)^{1/2} \langle \Delta \omega(q) \rangle \left\{ 
        \frac{1}{2} - \frac{\exp\left[-2\left(\omega_0/\Delta \omega \right)^2 \right]}{\pi \left[1 + \mathrm{erf}\left(\omega_0 / \Delta \omega \right) \right]^2} \right. \nonumber \\
         & \left. - \frac{\omega_0}{\Delta \omega} \frac{\exp\left[-\left(\omega_0/\Delta \omega \right)^2 \right]}{\sqrt{\pi} \left[1 + \mathrm{erf}\left(\omega_0 / \Delta \omega \right) \right]}
    \right\}^{-1/2}.
\end{align}
For positive-valued $\omega_0$ and $\Delta \omega$, this function has extreme limits
\begin{align}
    \lim_{\omega_0/\Delta \omega \to 0} \mathrm{FWHM} &= 2\left[\frac{2\pi \ln 2}{\pi - 2}\right]^{1/2} \langle \Delta \omega(q) \rangle \\
        & \approx (3.906)\langle \Delta \omega(q) \rangle, \nonumber \\
    \lim_{\omega_0/\Delta \omega \to \infty } \mathrm{FWHM} &= 2(2\ln 2)^{1/2} \langle \Delta \omega(q) \rangle \\
        & \approx (2.355)\langle \Delta \omega(q) \rangle. \nonumber
\end{align}

\end{document}